Martin Henry Higgins

Locational Marginal Pricing:

Towards a True Free Market in Power

Summary

*Nothing has done more to empower the free market, enterprise, and meritocracy than the spread of eletricity and power to everyone. The power system has been the precursor to the greatest period of innovation in our history and has meant that visionaries with revolutionary ideas can compete with those with capital, political power, and means. Electricity, therefore, has been the great equalising force of the last 150 years, enhancing the productivity of the masses and granting prosperity to whole swathes of our nation.*

*Whilst electricity has been one of the single largest innovations in enhancing the power of free markets, it is somewhat ironic that the way power is sold to consumers is largely unfree. The market is highly regulated, centralised, and is often used for political football by cynical politicians on both sides of the political spectrum.*

*Introducing Locational Marginal Pricing (LMP) into the UK grid system will increase economic freedom in the consumer markets for power, reduce prices for the poorest in the UK, decrease transmission losses, increase the permeation of low carbon generation in the grid, and incentivise invesment in the UK's Northern Powerhouse initiative.[1] LMP will not require cables to be dug up, nor useful assets discarded, but merely the implementation of a freer market in power pricing. The costs of the changeover will be minor and the technology required to make the change well tested and mature. LMP represents a fairer, less wasteful, and economically more sensible form of pricing policy that will enhance the economic freedom of those living in the British Isles.*

Introduction

The advent of electronics, electricity, and distributed communications has entailed the introduction of disruptive technologies in almost every part of our lives. Our ability to communicate, live, and work has been significantly enhanced by technology. In 1920, only 6% of homes in the UK had an electricity supply. The market was fragmented with no centralised infrastructure, standards, or regulation. By 1933, only thirteen years later, two-thirds of UK households had electricity in their homes.[2]

---
[1] https://northernpowerhouse.gov.uk/ (accessed 14 August 2019)

Stanley Baldwin, the Conservative PM at the time, recognised the importance of power in uplifting lives and growing the country. In 1926, Baldwin established the Electricity Supply Act, linking 122 of the most efficient power plants and establishing the National Grid. As grids grew, they became more reliable, so from a power system engineering perspective, centralisation and connection made perfect sense. Where previously, the UK had a series of fragmented electricity systems, now (and to this day) it had a centralised pan-UK system. The grid was strengthened through linkage, blackouts were reduced, and the reliability of the system increased. Economies-of-scale were also achieved in power plant construction, with bigger more efficient power replacing the smaller less efficient generating units. Baldwin's aims were fundamentally noble and his policies were reasonable, given the technology of the time and the need to rapidly electrify Britain. This urgency justified the centralisation of the grid but Baldwin could have never foreseen the smart grid technologies that are available today and may necessitate a different policy arrangement.

The government established the National Grid in tandem with a national price of energy for consumers. To this day, the national price for consumer power in the UK remains an unquestioned fact of everyday life. Consumers pay roughly the same price for power across the UK despite large differences in the price of geographic consumption. This proposal will argue that this disparity needs to addressed.

This report is organised as follows: First, the general principle behind LMP and how it differs from the current pricing system is outlined. Second, the impact of this model on consumer and prosumer activity, and how it might address some high-profile market issues is discussed. Finally, some case studies that show how individuals in the UK would be affected by LMP implementation are explored.

## A True Free Market for Power

LMP would create a location specific pricing model and tie the consumer price of power to its relative geographic cost of consumption. In essence, the idea is relatively simple. Whereas currently, the consumer is exposed to a fixed tariff price agreed with the utility provider, under LMP, areas where it is cheap for the market to consume would have a lower price of electricity and areas of high cost would have a higher price.

---

[2] https://www.bbc.co.uk/news/uk-politics-11619751 (accessed 14 August 2019)

Hidden from most consumers is the fact that different geographic areas have significant variations in the cost of consumption of electricity. Inefficiency losses of around 8-10% occur in the transportation of eletricity.[3] There are also limits to how much power lines can transmit which can mean that often more expensive, local generation is used.[4]

In practical terms, this means that areas of high demand that are far from generation assets have higher costs of consumption than areas adjacent to generation. A wind farm in Scotland can produce power for close to nothing but transmission costs and limits to transmission capacity mean that often their production is curtailed. To make up for the shortfall of generation, localised (to the demand) and more costly production, such as gas turbines or coal fire power stations are used.[5] LMP would mean charging individual consumers based on their respective *actual* cost of consumption rather than the current model of a fixed, national price charged to all consumers across the UK. LMP would recognise that the localised cost of power consumption varies greatly based on promixity to generation, network constrainment, and current demand, and would reflect this in pricing electricity.

### LMP Improving Consumer Markets

Under the current policy some parts of the UK are effectively subsidised via the centralised pricing model. Consumers living in areas of high demand and limited generation capacity, such as London and the South East, benefit from a blended price which does not reflect their higher cost of consumption. On the other hand, regions with significant local, low-cost generation, such as Scotland and the North East, effectively pay a premium on energy despite having costs of production that are close to zero for large portions of time. This is particulary egregious when you consider the GDP disparity of up to 2.5 times between these regions, meaning broadly speaking, the richer regions of the UK are subsidised by the poor when it comes to power. [6]

---

[3] https://assets.publishing.service.gov.uk/media/54eb5da5ed915d0cf7000010/Locational_pricing.pdf (accessed 14 August 2019)
[4] Transmission networks are limited in how much power they can transmit by the size and number of overhead cables. Continually breaking these limits significantly reduces asset lifetimes.
[5] Wind farms and nuclear produce power for a price of close to free, the high the % nuclear and wind in our energy mix the cheaper our cost of power per kW on average.

To implement this plan, every energy consumer would have a smart meter installed in their home, which would record the rate, price-level, and the time of consumption. The meter would have a two-way-feed to inform the consumer of the current price of power within their area at that given period of time. The price would use the consumer's post code or GPS tracking to communicate a location adjusted price.

For those in power poverty, around 23% of their income is spent on power and fuel.[7] The poorest users of eletricity, those currently on key-meters, may benefit the most from an LMP pricing policy.[8] A time-based LMP model would mean consumption at times and areas of low demand would be significantly cheaper, particulary in those regions of high generation. Individuals could make decisions about consumption, choosing to heat homes or cook when power is close to free in price and opting for lower consumption during peak pricing hours. For those people in energy poverty who are income-poor but time-rich, such as pensioners or the unemployed, this policy will give them an option beyond simply reducing consumption. By amending the way they use power, they could get their full usage at a significantly reduced cost and have more discretionary income.

By disincentivising peak consumption, demand at peak times will be reduced through consumer choice. This will result in several secondary benefits. One benefit is that network traffic will fall, resulting in lower grid constrainment. This in turn would mean cheaper generating assets can be used as a greater percentage of our energy mix as peak power consumption drops. In the long-run, this would mean a reduction in the overall cost of energy across the whole of the UK. The cheaper forms of production, usually wind, solar, and nuclear, are also generally low carbon resources. This means that by reducing grid congestion through LMP we also contribute in reducing $CO_2$ outputs, thereby helping to meet our climate change commitments.

---

[6] https://www.oecd.org/cfe/UNITED-KINGDOM-Regions-and-Cities-2018.pdf (accessed 14 August 2019)
[7] https://assets.publishing.service.gov.uk/media/54eb5da5ed915d0cf7000010/Locational_pricing.pdf (accessed 14 August 2019)
[8] Old coal towns such as those in The Valleys and Midlands would have particularly low prices. This is because in the past, generation stations were built close to coal fields to minimise the distance coal had to travel before reaching the furnace.

## Fixing Broken Prosumer Markets with LMP

Single price, wholesale policies are often applied to consumers who also produce power (prosumers) i.e. individuals with solar installations on their homes. The feed-in tarifff (FIT) is a common policy mechanism that is employed across many European states whereby a single (usually subsidised) price is offered to consumers willing to feed power back into the grid using low carbon generation. Naturally, for most prosumers, solar is the only viable form of prosumer activity available due to the large scale investments and expertise required in wind or other forms of generation. Solar panels, however, are an uncontrollable and unpredictable power source that can create more issues for power system operators than they solve. Due to the decentralised nature of prosumer activity, the level of generation is usually unknown by the power system operators and is often modelled by power system operators as a negative load (something that can't be controlled or counted on for capacity) rather than actual generation.[9] Whilst the implementation of FITs can be politically necessary to create the illusion that a governement is aiming to reduce its carbon footprint, often solar friendly policies can do the opposite. Germany offered generous FITs (as high as 50 c/KWh vs a national average price of 7 c/KWh) resulting in massive amounts of investment in consumer solar panels.[10] Consequently, Germany now suffers frequently with issues of overvoltage due to excessive and unecessary solar production and at certain times will charge generators to export to the grid in order to try and prevent damage to the system (also known as negative prices). Surprisingly, Germany's carbon footprint has actually increased from 2009-2018 as inflexibility in production due to solar panels has meant a further reliance on coal.[11] California also has issues caused by excessive solar generation and even has to pay other states to take its power surplus, resulting in some of the highest consumer prices for power in the United States.[12] Large power surpluses are problematic for power systems as they force transmission assets past their heat and voltage limits. This can destroy systems over time and cause long blackouts as replacing assets is difficult and costly.

---

[9] As of 2017 National Grid believes they have 12.8GW of solar active on the grid but is unsure. Solar panels also do not provide inertial response and create backwards power flow in distribution systems which have create innumerable challenges for distribution operators.
[10] https://www.netztransparenz.de/EEG/Verguetungs-und-Umlagekategorien (accessed 14 August 2019)
[11] https://www.cleanenergywire.org/factsheets/germanys-greenhouse-gas-emissions-and-climate-targets (accessed 14 August 2019)
[12] https://www.bbc.co.uk/news/business-40434392 (accessed 14 August 2019)

Contrary to what these cases indicate, solar electricty has a place in the future of smart electricity systems. The issues discussed in the previous paragraph have occurred because the single pricing models encouraged consumers to build panels in areas of little demand where there was no market or use for the power. The generous subsidies helped compound this issue, encouraging even more investment in unecessary and useless solar panels. However, if the FITs had been offered on a geographic basis and only in areas of high demand, these issues would not have occurred as frequently. Additionally, widescale FITs like the ones used in Germany can create a direct counter-incentive to good prosumer activity. Storage units such as batteries are a great solution from a grid perspective as they can take pressure off the power system by exporting during peak consumption or charging during peak production. However, prosumers were directly dis-incentivised by the high FIT rates as the price of exporting to the grid was significantly higher than directly consuming. This meant it was often better for a prosumer to export to the grid rather than use their own power supply. The German approach was an abmismal failure in power policy that has cost taxpayers billions and has done little to reduce carbon emissions.

An LMP pricing model applied to prosumer activity would have prevented these issues from occuring. The localised price incentive would mean prosumers would have an incentive to produce energy only in those areas that had a market for the power. Areas of high generation and low demand would offer lower prices, reducing the incentive to build more panels in these regions. Conversely, regions of high demand would have larger incentives to build panels as compensation rates for prosumers increase. Governments, cognisant of public relationships and their obligation to be "seen doing something" on climate change could still offer subsidises. However, under LMP they would simply be offered on a geographic basis to match market demand.

Incorporating a basic time-based pricing model would greatly incentivise the building of storage devices such as batteries and take more pressure off the grid. Prosumers would be incentivised to consume their own supply via the storage device when power was expensive or export to the grid when local demand was high.

These changes to the prosumer markets can be implemented easily with smart metering devices and a post code based, power pricing system. Prosumer LMP will result in a more appropriate allocation of assets, less money spent on worthless generation and a more reslient system. Subsidies can still exist but would be applied intelligently. Decarbonisation would continue at similar rates, but at a significantly reduced cost.

## Case Studies

This section will consider some case studies outlining how LMP might impact individuals within the UK.

### Case Study 1 - South Coast Retiree - Living Adjacent to a Proposed Wind Farm

Consider the case of a mid to high-asset value but low-income individual living in a highly populated area of the country. The majority of their wealth is locked up in the family home and the region where they live has a high demand for power coupled with only limited generation. Currently, the local region relies on a combination of low-cost power imported from the Midlands and higher cost local generators which turn on during periods of peak demand.

Now consider that a wind farm is proposed to be built adajcent to the Retiree's home. How are they incentivised to act from a market perspective? Under the current scenario, the Retiree gains nothing from the building of the wind farm but may experience several disadvantages. It is common knowledge that some individuals consider wind farms unattractive. The wind farm may directly impact their home's asset value. It will therefore be in the direct interest of the Retiree to oppose and frustrate the building of the farm (even if they don't particulary mind or quite like wind farms on a personal level). This phenemonem, often known as "not-in-my-backyard" (Nimbyism), is a perfectly rational response for most consumers who are offered several disadvantages and limited benefits to allowing construction near their homes. Nimbyism can cause lawsuits, production hold-ups, and often result in sub-optimal placement of resources. One obvious example of this is how the government now opts for offshore wind farms over onshore as these do not suffer from Nimbyism. Off-shore farms, however, cost about 50% more to build per MW of capacity, which directly increases costs for all consumers of power.[13]

LMP addresses these issues. Under LMP pricing, the Retiree would be offered a direct incentive to support the building of the wind farm. The construction of the farm would mean energy supply would be taken locally, reducing losses and utilising a cheaper source. It would also reduce constrainment, meaning less reliance on the expensive forms of local generation. The incentives between the consumer and producer are aligned more closely under the free market LMP system. There are also significant secondary benefits in bureacracy and cost, including less time spent on planning permission lawsuits, less bureacracy, and lower construction times before the assets become revenue generating.

---

[13] https://www.irena.org/documentdownloads/publications/re_technologies_cost_analysis-wind_power.pdf (accessed 14 August 2019)

In short, LMP would mean that the Retiree is compensated directly for their potential loss of value to their assets. LMP addresses Nimbyism by offering counter-incentives to the perceived loss of value from the building of generating sites. LMP would mean more generation assets built closer to demand, resulting in lower costs, lower losses, less constrainment, and more low carbon resources. It would also confer secondary benefits through reductions in bureacracy, quicker building, and lower construction costs.

### Case Study 2 – European Manufacturer - Selecting a Factory Location

Now, consider the case of a manufacturer of complex machined goods looking to set up somewhere in Europe. The manufacturing process is highly automated with the chief variable cost being electrical power. Currently, the UK's consumer price of electricity is one of the highest in Europe with 32 nations on the European sub-continent offering cheaper consumer prices.[14] The UK offers excellent business protections, favourable employee hiring/firing practices and a legal system that makes it a highly attractive region for foreign direct investment. However, under the current scenario of a national consumer price, the manufacturer is unlikely to choose the UK as its production base given the multiple low-cost alternatives spread across Europe.

LMP would remedy this issue. It would create a geographic pricing model within the UK that would offer significant pricing variation across the different geographic regions. The manufacturer would have the choice to invest either in high cost regions, such as London where high demand and low generation drive up prices, or lower cost areas, such as the Midlands or North East which, due to being historic coal towns, have large generation facilities. There could be as much variation of consumer prices within the UK as across the whole of the EU giving real choice to investors. Some regions would offer lower prices, which would incentivise building high-power consumption facilities in these regions, and others, such as London, would not. In this way, LMP would directly benefit disadvantaged UK regions with increased productivity and employment opportunities. LMP would also mean greater investment in former coal-mining towns and would further enfranchise the government's Northern Power House strategy. Scotland would also likely benefit with its high concentration of cheap wind power.

---

[14] https://ec.europa.eu/eurostat/statistics-explained/index.php/Electricity_price_statistics (accessed 14 August 2019)

### Case Study 3 – Islington Based Prosumer – Considering Solar Panels

In the short term, some individuals may have increased prices within their local region but in the long term it is expected that LMP would reduce power prices acoss the board. Consider an individual based in a high demand, low-generation area with means to invest in prosumer activity. Currently it takes twelve years for prosumer investment in solar panels to repay the intial capital investment.[15] For most individuals, this represents a marginal gain at best, as the long payback rate also comes with significant outlay, limited medium term cash-out ability, and risk associated with loss and malfunction. In fact, it seems reasonable that, at this rate of return, a significant portion of the investment into solar will be motivated by personal beliefs on climate change rather than investment incentives.

However, under LMP a potential prosumer might be highly incentivised to participate in the prosumer market, irrespective of his personal opinions. An individual in an area such as Islington would see the payback periods for solar investment drop significantly as his consumptive power costs increase. He would also be highly incentivised to invest in an accompanying storage system to allow him to use as much of his own power as possible. In turn, increased investment at the local level, close to demand, will mean prices drop for other consumers in the region as prosumers rise to match the local demand. This would also increase the UK's low carbon generation as a percentage of its energy mix whilst also protecting the grid and market from excessive uncontrollable generation.

---

[15] https://www.renewableenergyhub.co.uk/main/solar-panels/return-on-investment-for-pv-return-on-investment-for-solar-panels-roi-for-pv-solar-panels-pay-back/ (accessed 14 August 2019)

## LMP Towards Economic Freedom

Whilst the free market system we enjoy in the West has been greatly enhanced by the modern power system, the power system itself has been largely insulated from the benefits of a free market economy. This insulation has had serious effects, demonstrated by the flawed prosumer markets in Germany and California, which have failed in their aims of reducing carbon emissions at the cost of billions to the tax payer.

Through a geographic and time-based pricing system, consumers can make their own decisions on the generation and consumption of power, uninsulated from the consequences of their choices, resulting in a cheaper, more reliable, and fairer power pricing strategy. LMP would address issues such as nimbyism, prosumer over-supply, and power poverty in the UK. In the short-run, the strategy would benefit those in the poorest regions of the UK, particulary the North, Midlands, and Welsh Valley regions. In the long-run, the whole of the UK would benefit from cheaper prices, reduction in waste, and a more secure decentralised power system. LMP will offer a lifeline to the poorest in the UK and increase the UK's competitiveness as a manufacturing base. By offering a geographic pricing model some regions of the UK would become targets for new factories and investment in modern industries whilst consumers in these regions would also benefit from lower cost and more flexibility in their power consumption.

When it comes to electricity, price control is the most frequently uttered sentence by politicians seeking to secure votes for the next election. However, when consumers vote with their wallets and feet, they always opt for the freer market where decision-making is devolved. LMP would give politicians a solution to problems of energy poverty that does not simply throw money at the problem, which does not reduce economic freedom, and will not endanger the UK's power supply. The changes required are modest, the technology is mature and the time is right to finally implement LMP in the UK power system.